\documentclass{article}
\usepackage[utf8]{inputenc}
\usepackage{authblk}
\usepackage[dvips]{graphicx}
\graphicspath{{noiseimages/}}
\usepackage{xcolor}
\usepackage[T2A]{fontenc}
\usepackage{tikz}
\usepackage{MnSymbol}
\usepackage{pgfplots}
\usepackage{mathrsfs} 
\pgfplotsset{compat=1.7}
\usetikzlibrary{intersections, pgfplots.fillbetween}
\usetikzlibrary{decorations.pathmorphing}
\usetikzlibrary{arrows.meta}
\usepackage[mode=buildnew]{standalone}
\usepackage[toc,page]{appendix}
\usepackage{cite}
\usepackage{amsmath}
\usepackage{float}
 \usepackage{bm}
\usepackage{a4wide}
\usepackage{bm}

\usepackage[english]{babel}
\usepackage{hyperref}
\definecolor{urlcolor}{HTML}{990000}
\definecolor{linkcolor}{HTML}{005F5F} 
\hypersetup{pdfstartview=FitH,  linkcolor=linkcolor,urlcolor=urlcolor, colorlinks=true,citecolor=blue}
\usepackage[page]{appendix}

\author[1,2]{D.V.Diakonov\footnote{\tt dmitrii.dyakonov@phystech.edu}}

\affil[1]{\itshape Institutskii per, 9, Moscow Institute of Physics and Technology, 141700, Dolgoprudny, Russia}
\affil[2]{\itshape Bol'shoi Karetnyi per., 19, Institute for Information Transmission Problems, 127994, Moscow, Russia}

\date{}

\title{\textcolor{black}{De Sitter entropy: on-shell versus off-shell}}

\begin{document}

\numberwithin{equation}{section}

\maketitle

\begin{abstract}
Attributing thermodynamic properties to the Bunch-Davies state in static patch of de Sitter space and setting the corresponding equations of state, we demonstrate that, for pure gravity, the bulk entropy — computed on-shell as a volume integral in de Sitter space — coincides with the Wald entropy (area law) in any spacetime dimension and for any theory of \(f(R)\) gravity. We extend this result to the renormalized entanglement entropy of a non-minimally coupled scalar field. From the on-shell perspective, entropy emerges as a bulk contribution, whereas from the off-shell viewpoint, it manifests as a boundary (horizon) contribution. As a result, in de Sitter space, generalized entropy can be understood in two distinct ways: either as a bulk or as a boundary contribution.
\end{abstract}
\newpage

\tableofcontents

\section{Introduction}
Space-times with Killing horizons behave like thermodynamic systems, characterized by temperature and entropy \cite{Bekenstein:1973ur,Hawking:1975vcx}. Some progress has been made in understanding the thermodynamic properties of black holes and the statistical mechanical interpretation of entropy. In general, it is assumed that since a black hole is a compact object, we can count the microstates that constitute it and are hidden behind the horizon; for more details, see the review \cite{Carlip:2014pma}.

The Bekenstein-Hawking entropy $S_A(\text{gravity})=\frac{A}{4}$ is proportional to the area of the Killing horizon, which makes it significantly different from the ordinary entropy of a thermal gas, which is proportional to the volume of space. If we take into account matter, we should compute the generalized entropy $S_{\text{gen}}=S_A(\text{gravity})+S_A(\text{matter})$, where $S_A(\text{matter})$ is the entanglement entropy, which has an ultraviolet divergence proportional to the horizon area. Such terms should be properly renormalized by redefinition of coupling constants in $S_A(\text{gravity})$. The area law of the entanglement entropy is explained by observing that, due to short-distance correlations, only modes located in a small vicinity close to the surface, separeting two regions of space contribute to the entropy \cite{Srednicki:1993im,Frolov:1993ym}. Hence, for space-times with a Killing horizons, it is assumed that the generalized entropy satisfies the area law.

There is a number of differences between cosmological and black hole horizons. The cosmological horizon is observer-dependent. Moreover, an observer cannot approach or cross their own cosmological horizons. This leads to difficulties in providing a statistical mechanical interpretation of the entropy in terms of microstates located behind the cosmological horizons. If there are microstates that we cannot observe, then from the perspective of another observer, those states are a part of their accessible region. Therefore, there should be a consistent picture across different observational frameworks. Hence, there is a question of how we should understand the area law for the entropy in de Sitter space-time.

In this note, we show that instead of interpreting generalized de Sitter entropy as an area law, it can be interpreted as a bulk contribution to the entropy\footnote{We are discussing here only de Sitter invarian Bunch-Davies state in static patch of de Sitter space}. This bulk contribution precisely follows the area law. Hence, in de Sitter space, there are two ways of thinking about entropy: as a bulk or as a boundary (horizon) contribution. This implies that, since the interpretation of de Sitter entropy in terms of microstates located behind the cosmological horizon is heavily constrained, alternative explanations can be thought in the bulk space as some property of the Bunch-Davies state. 

The main result of this paper is that bulk entropy, computed on-shell in de Sitter space for any space dimension and any theory of \(f(R)\) gravity, coincides with the Wald entropy \cite{Wald:1993nt}:
\begin{align}
    S^{\text{on shell}}_V(\text{gravity})=S^{\text{off shell}}_A(\text{gravity})=\frac{A}{4}f'(R),
\end{align}
and we generalize this statement for the entanglement entropy of a non-minimally coupled scalar field:
\begin{align}
    S^{\text{on shell}}_V(\text{matter})=S^{\text{off shell}}_A(\text{matter}).
\end{align}
The bulk entropy $S^{\text{on shell}}_V$ is computed on-shell, i.e., by attributing thermodynamic properties to the background Fock space ground state (Bunch-Davies state) and setting the equations of state $\rho =-p=\frac{\Lambda(T_H)}{8\pi}$ and $V\sim\frac{1}{H^{d-1}}\sim 1/T_H^{d-1}$, where $d$ is dimention of de Sitter space, we compute the entropy as a volume integral, rather than as a contribution from the horizon, varying only the Gibbons-Hawking temperature\footnote{The origin of this equation is explained below.}:
\begin{align}
\label{S_Vwithminussign}
    S^\text{on shell}_V = -\int \frac{\delta Q}{T_H} = -\int_{0}^{T_H} \frac{d T_H}{T_H}  \int_{r_{dS} =\frac{1}{2 \pi T_H}} d^{d-1} x \sqrt{g}\ \frac{\partial \rho(T_H) }{\partial T_H}.
\end{align}
We use the first law of thermodynamics with the minus sign, because de Sitter entropy decrease if we add small amount of energy. The same opposite sign also arises if we add a black hole in the de Sitter space \cite{Gibbons:1976ue}, and it is one of the puzzling questions of how the de Sitter thermodynamics should be understood \cite{Banihashemi:2022htw}.
 
The area law $S^{\text{off shell}}_A$ is computed off-shell. Namely, we fix the Gibbons-Hawking temperature, but do not fix the period of Euclidean time $\tau=\tau+\beta$, so that $\alpha=\beta/\beta_H \ne1$. Then the Euclidean de Sitter space $\mathcal{M}_\alpha$ becomes a manifold with conical singularities on the horizon surface. Since all curvature tensors contain delta-like singularities on the horizon surface \cite{Fursaev:1995ef} (for example: $R_{\mathcal{M}_\alpha}=R_{\mathcal{M}}+(1-\alpha)\delta(\Sigma)+...$), we can expand the Euclidean effective action close to the Gibbons-Hawking temperature and compute the entropy:
\begin{align}
    S^{\text{off shell}}_A=-(\alpha\partial_\alpha-1)\log Z\Bigg|_{\alpha=1}\sim A, 
\end{align}
showing that the off-shell method produces the area law for the entropy.

The paper is organized as follows: In Sec. \ref{2}, we describe the geometry of de Sitter space. In Sec. \ref{3}, we provide an overview of different methods for deriving the temperature and entropy in de Sitter space. In Sec. \ref{4}, we describe the method of computing bulk entropy using the on-shell approach. In Sec. \ref{5}, we obtain the bulk entropy for the scalar field theory. In Sec. \ref{6}, we discuss why the two methods yield the same result and identify a subtle issue in our proof. Finally, in Sec. \ref{7}, we explicitly derive the renormalized entanglement entropy.

\section{Geometry of de Sitter space \label{2}}
De Sitter space is the vacuum solution of the Einstein equations:
\begin{align}
    R_{\mu\nu} - \frac{1}{2} R g_{\mu \nu} + \Lambda g_{\mu\nu} = 0,
\end{align}
where: 
\begin{align}
\label{R ds}
    R_{\mu \nu} = (d-1)H^2 g_{\mu\nu}, \quad R = d (d-1) H^2, \quad  \Lambda = \frac{(d-1)(d-2)}{2} H^2,
\end{align}
and \( H \) is the Hubble constant, $d$ is the dimention of space-time. Here we ussume signature $(-,+,...,+)$ and adopt Planck units: $\hbar = c = G = k_B = 1$. But below everywhere we assume analytical continuation to the sphere, which is valid for Bunch-Davies state.

The \( d \)-dimensional de Sitter space-time can be visualized as a one-sheeted hyperboloid embedded in a \( d+1 \)-dimensional ambient Minkowski space-time, described by:
\begin{align} 
\label{dssa}
dS_d = \{ X \in \mathbf{R}^{d+1}, \ \  X^\alpha X_\alpha = -X_{0}^2 + \sum_i X_{i}^2 = H^{-2}\}.
\end{align}

In addition, de Sitter space serves as a solution of modified theories of gravity. To illustrate this, consider \( f(R) \) gravity, which is described by the action:
\begin{align}
\label{action f(R)}
    W = \frac{1}{16 \pi} \int d^{d} x \sqrt{g} \left( f(R) - 2\Lambda \right) + W_{\text{matter}}.
\end{align}
The variation of the action \eqref{action f(R)} with respect to the metric yields \cite{Sotiriou:2008rp}:
\begin{align}
\label{eq motion f(R)}
     f'(R) R_{\mu\nu} - \frac{1}{2} f(R) g_{\mu\nu} + \left( g_{\mu\nu} \Box - \nabla_\mu \nabla_\nu \right) f'(R) + \Lambda g_{\mu \nu} = 8 \pi T_{\mu\nu}^{\text{matter}}.
\end{align}
Thus, de Sitter space, with the Hubble constant \( H \), is the solution to the field equations in the absence of matter and the cosmological constant is given by:
\begin{align}
\label{Lambda}
    \Lambda =  \left(\frac{1}{2} f(R) -\frac{1}{d} f'(R) R \right) \Bigg|_{R = (d-1)d H^2}.
\end{align}
Therefore, de Sitter space is the solution in a wide range of gravitational theories. From a geometric perspective, these spaces may be considered equivalent, but, as we will discuss later, they exhibit different values for the vacuum energy and entropy.

The static coordinates of de Sitter space are defined by:
\begin{equation}
\label{coordinates}
 X = \begin{cases}
   X^0 = H^{-1} \sqrt{1 - r^2 H^2} \sinh(t H), \\
   X^i = r z_i, \quad \quad i = 1, \ldots, d-1, \\
   X^d = \pm H^{-1} \sqrt{1 - r^2 H^2} \cosh(t H),
 \end{cases}
 \qquad t \in (-\infty, \infty), \ r \in (0, H^{-1}),
\end{equation}
where \( z_i \) are the coordinates on the \( (d-2) \)-dimensional sphere, and the \( \pm \) in \( X^d \) denotes the right or left static de Sitter patch. In these coordinates, the de Sitter metric, induced from the ambient space-time, takes the form:
\begin{align}
\label{metric}
    ds^2 = -(1 - r^2 H^2) dt^2 + \frac{dr^2}{1 - r^2 H^2} + r^2 d\Omega_{d-2}^2.
\end{align}
The static coordinates are bounded by the cosmological horizon:
\begin{align}
r_{\text{horizon}} = \frac{1}{H},    
\end{align}
where the metric degenerates. The volume of the static patch and the area of the cosmological horizon are finite:
\begin{align}
\label{volum and area}
   V = \frac{\pi^{\frac{d-1}{2}}}{\Gamma\left(\frac{d+1}{2}\right)} \frac{1}{H^{d-1}}, \quad \text{and} \quad A = \frac{2\pi^{\frac{d-1}{2}}}{\Gamma\left(\frac{d-1}{2}\right)} \frac{1}{H^{d-2}}.
\end{align}

The geodesic distance \( \sigma \) in de Sitter space is given by the de Sitter-invariant scalar product:
\begin{gather}
\label{geod dist}
    Z_{12} = \cosh(H \sigma) = \frac{X_1^{\alpha} X_{2\ \alpha}}{H^{-2}} = \sqrt{1 - r_1^2 H^2} \sqrt{1 - r_2^2 H^2} \cosh\left(H(t_1 - t_2)\right) + H^2 r_1 r_2 z^1_i z^2_i.
\end{gather}
\section{Thermodynamics of de Sitter space \label{3}}
In this section, we briefly list the methods, without providing detailed calculations, which demonstrate that an observer in the static patch of de Sitter space sees isotropic radiation with the Gibbons-Hawking temperature \cite{Gibbons:1976ue,Gibbons:1977mu}:
\begin{align}
\label{T hawking}
    T_H=\frac{H}{2\pi}.
\end{align}
We also review some methods that show that, in the semiclassical approximation, the entropy in de Sitter space obeys the area law \cite{Gibbons:1976ue,Gibbons:1977mu}:
\begin{align}
    S_A=\frac{A}{4}.
\end{align}
Modifications to the area law that arise when quantum corrections are taken into account will be considered in the next section.

The Gibbons-Hawking temperature cannot be observed in practice, as this temperature in our Universe is $10^{-30} K$, which is much lower than the temperature of the cosmic microwave background, $T=2.73 K$. However, this temperature implies the existence of the entropy in de Sitter space, which is given by $2.6 \cdot 10^{122}$, vastly exceeding the entropy of all the matter and energy in our Universe, which is on the order of $ 10^{104}$ \cite{Egan:2009yy}.

\subsection{The de Sitter temperature}
Similar to black hole thermodynamics \cite{Carlip:2014pma}, there are several heuristic methods to predict the presence of the Gibbons-Hawking temperature \eqref{T hawking}.

\begin{enumerate}
    \item \textbf{Imaginary time periodicity trick} \\
Let us consider the line element \eqref{metric} of the static de Sitter patch in Euclidean signature by analytically continuing \(t \to -i\tau\). Then, the line element near the cosmological horizon \((r \approx H^{-1} - H \epsilon^2/2)\) is given by: 
\begin{align}
    ds^2 \approx d\epsilon^2 + \epsilon^2 d(H \tau)^2 + H^{-2} d\Omega^2.
\end{align}
To avoid a conical singularity in the Euclidean metric, we must impose the periodic condition \(\tau \sim \tau + \frac{2\pi}{H}\). This requirement leads to the periodicity of the correlation functions in Euclidean time \(\tau\), with the period equal to the inverse temperature as required in thermal field theory, by the Kubo-Martin-Schwinger condition \cite{Kubo:1957mj,Martin:1959jp}. Therefore, we conclude that the temperature is given by \eqref{T hawking}.

    \item \textbf{Tunneling exponent} \\
For a massless particle, the Hamilton-Jacobi equation is given by \(g^{\mu \nu} \partial_\mu W \partial_\nu W = 0\), where \(W\) is the classical action of the particle. If the classical action has an imaginary part for some particle trajectory, this implies that the gravitational background in question is unstable with respect to radiation, and the emission rate is given by \cite{Akhmedov:2006un,Akhmedov:2006pg}:
\begin{align}
    \Gamma \sim \left| e^{i W} \right|^2 = e^{-2 \text{Im} W},
\end{align}
which describes the tunneling of the particle through the gravitational barrier (horizon). We can split the action into the time and spatial parts: \(W = E t - \int p_r dr\), where \(E\) is the particle energy and \(p_r = \frac{E}{1 - r^2 H^2}\) is the canonical momentum. Integration around the pole \(r = H^{-1}\) gives an imaginary contribution from the spatial part: \(\text{Im} \int p_r dr = \pm E \frac{\pi}{2H}\), where the \(+\) sign is for the ingoing particle and the \(-\) sign is for the outgoing particle. As shown in \cite{Akhmedov:2008ru}, we must also take into account the imaginary contribution from the temporal part in $W$. Since the particle can move away along a classically allowed trajectory, which should yield the vanishing imaginary part of the classical action, we can find that \(\text{Im}(E t) = -E \frac{\pi}{2H}\). Thus, the classical action for an emitted particle from the cosmological horizon contributes an imaginary part, and the emission rate takes the form \cite{Sekiwa:2008gk}:
\begin{align}
    \Gamma \sim e^{-\frac{2\pi}{H} E},
\end{align}
which yields the Gibbons-Hawking temperature \eqref{T hawking}.

    \item \textbf{KMS condition} \\
In de Sitter spacetime, the maximally symmetric propagator that satisfies the correct Hadamard condition is the Bunch-Davies Green function \cite{Bunch:1978yq}. This propagetor depends on the geodesic distance \eqref{geod dist}, which is periodic in imaginary time with the period equal to the inverse Gibbons-Hawking temperature \eqref{T hawking}. Hence, the Green function obeys the Kubo-Martin-Schwinger condition \cite{Kubo:1957mj,Martin:1959jp}. Moreover, one can explicitly show that the Bunch-Davies Green function is equal to a thermal propagator in the static patch of the de Sitter space \cite{Akhmedov:2020qxd}: 
\begin{align}
\label{Green BD}
    G_{BD}(Z_{1,2})=\frac{1}{Z} Tr\left[e^{-\beta_H H} \phi(x_1)\phi(x_2)\right],
\end{align}
where $\beta_H$ is the inverse Gibbons-Hawking temperature \eqref{T hawking}. Thus, observers in de Sitter space sees the Bunch-Davies vacuum as a thermal bath.

    \item \textbf{Reduced density matrix} \\
Another method to show that the Bunch-Davies state in the static patch corresponds to a thermal density matrix with the Gibbons-Hawking temperature \eqref{T hawking} is by considering the reduced density matrix \cite{Bousso:2001mw}. Let us express the Bunch-Davies state as a superposition of states in the right and left de Sitter patches \eqref{coordinates}:
\begin{align}
    |BD\rangle =\prod_{\omega} \left(1-e^{-\frac{2\pi}{H} \omega}\right) \sum_{n_{\omega}} e^{-\frac{\pi}{H} \omega n_\omega} |n_\omega,L\rangle \otimes |n_\omega,R\rangle.
\end{align}
The observer moving along a timelike geodesic in the right static de Sitter patch does not receive information from the left patch. Therefore, the reduced density matrix for the right patch is given by the partial trace \cite{Bousso:2001mw}: 
\begin{align}
    \hat{ \rho}_R = Tr_L|BD\rangle \langle BD| = \prod_\omega \left[\left(1-e^{-\frac{2\pi}{H} \omega}\right) \sum_{n_{\omega}} e^{-\frac{2\pi}{H} \omega n_\omega} |n_\omega,R\rangle \langle n_\omega,R | \right].
\end{align}
As a result, the reduced density matrix describes a thermal density matrix with the Gibbons-Hawking temperature \eqref{T hawking}.

    \item \textbf{Unruh-DeWitt detector} \\
A more intuitive way to demonstrate that observers see the Bunch-Davies state as a thermal bath is by using a detector ("thermometer") that interacts with the field and, after thermalization, measures the ambient temperature. The interaction between the detector and the field is described by the Hamiltonian \(H_{int}=g m(t) \phi(x(t))\), where \(x(t)\) is the observer’s worldline, and the Hamiltonian of the detector is \(H_{d}\), with the energy eigenstates is defined by \(H_d|E_i\rangle =E_i|E_i\rangle\). The probability per unit time that the system transitions from the initial state \(|0\rangle |E_i\rangle\) to a final state \(|\phi\rangle |E_j\rangle\) is given by \cite{Spradlin:2001pw,Bousso:2001mw}:
\begin{align}
    \dot{P}\left(E_i\to E_j\right) =g^2 |m_{ij}|^2 \int dt e^{-(E_j-E_i)t} G_{BD}(Z=\cosh(t)),
\end{align}
where we sum over all possible final states of the scalar field since the detector does not measure it. From the analytical properties of the Green function \eqref{Green BD}, we can show that the probabilities satisfy the relation: 
\begin{align}
    \dot{P}\left(E_i\to E_j\right)=\dot{P}\left(E_j\to E_i\right) e^{-\beta_H (E_j-E_i)}.
\end{align}
Thus, if the energy levels of the detector are thermally populated such that \(N_i=N e^{-\beta_H E_i}\), we can conclude that there is no change in the distribution: \(N_i \dot{P}\left(E_i\to E_j\right)-N_j \dot{P}\left(E_j\to E_i\right)=0\). Therefore, the detector is in thermal equilibrium, and free falling observers in de Sitter space measure the Gibbons-Hawking temperature. The method $1,3$ and $5$ are related to each other of course. 
\end{enumerate}

\subsection{The de Sitter entropy}

The entropy of the cosmological horizon is given by the Bekenstein-Hawking area formula \cite{Gibbons:1976ue,Gibbons:1977mu}, which is valid only in the context of Einstein gravity. For a general theory of \( f(R) \) gravity, the area formula must be modified and is expressed using the Wald relation \cite{Wald:1993nt,Solodukhin:2011gn}: 
\begin{align}
\label{entropy f(R) general}
    S_H = \frac{A}{4} f'(R).
\end{align}

In this section, we review several methods that demonstrate how one can obtain the area formula \eqref{entropy f(R) general} for the entropy in de Sitter space. In this paper we essentially assume that gravity theory in de Sitter space is obtained via analytical continuation from the sphere. Such an approach is of course arguable, but this discussion goes beyond the scope of our paper.  
\begin{enumerate}

\item \textbf{Replica trick} \\
Let us consider a general curved spacetime \( \mathcal{M} \) with a Killing horizon with the geometric temperature given by \( T_H=\kappa/(2\pi) \), where \( \kappa \) is the surface gravity, which we will normalize to one. In Euclidean time, we impose periodic boundary conditions \( \tau \sim \tau+\beta \), where the inverse temperature is \( \beta=2\pi \alpha \). Consequently, the Euclidean space exhibits a conical singularity, implying that the Euclidean geometry near the horizon is approximately \( \mathcal{M}_\alpha\approx \mathcal{C}_\alpha\times \Sigma \). In the limit when the deficite angle is set to zero, \( \alpha \to 1 \), the Riemann tensor contains delta-like surface contributions near the horizon \cite{Fursaev:1995ef,Solodukhin:2011gn}: 
    \begin{align}
          \lim_{\alpha \to 1}  R^{\mathcal{M}_\alpha}_{\mu\nu\rho\sigma}=R^{\mathcal{M}}_{\mu\nu\rho\sigma}+2\pi(1-\alpha)  \left(n_\mu^1 n_\rho^2-n_\mu^2 n_\rho^1\right)\left(n_\nu^1 n_\sigma^2-n_\nu^2 n_\sigma^1\right)  \delta\left(\Sigma\right)+O\left((1-\alpha)^2\right),
    \end{align}
where \( n^{1,2} \) are two orthonormal vectors, which are orthogonal to the horizon surface \( \Sigma \). The classical action in the limit \( \alpha \to 1 \) is given by:
    \begin{gather}
    \label{Wq}
        W^E [\alpha]
        = \\
        =\nonumber
        -\alpha \frac{1}{16 \pi}\int_\mathcal{M} d^{d}x \sqrt{g}  \left(f \left(R^{\mathcal{M}}\right)-2 \Lambda\right) - 
        \\ 
        \nonumber
        -
        (1-\alpha) \frac{1}{4}\int_\mathcal{M} d^{d}x \sqrt{g}  \delta\left(\Sigma\right)  f'\left(R^{\mathcal{M}}\right) +
        O\left((1-\alpha)^2\right),
    \end{gather}
where the negative sign in the action arises because, after analytical continuation \( t \to -i\tau \), we define the Euclidean effective action as \( e^{i W} \to e^{-W^E} \). Applying the replica formula, we find that only the surface term contributes to the entropy: 
\begin{align}
        S_A=(\alpha\partial_\alpha-1)W^E[\alpha]\Bigg|_{\alpha=1}= \frac{A}{4}  f'(R).
    \end{align}
\item \textbf{Free energy \( F=E-T_H S \)}\\
The partition function for thermal gravity can be evaluated in the semiclassical limit using the on-shell gravity action in Euclidean signature \cite{Galante:2023uyf,Gibbons:1976ue}: 
    \begin{align}
        Z=e^{-\beta F}=\int D g_{ab} e^{-W^E(g)}\approx e^{-W^E(g_{dS})}.
    \end{align}
Using standard expression for the free energy: \( F=E-T_H S \), where \( E \) is the energy (set to zero for empty de Sitter space), we can show that the entropy for the general theory of \( f(R) \) gravity is:
    \begin{align}
    \label{S wald from F=U-TS}
        S=-\beta F=-W^E(g_{dS})=\frac{1}{16 \pi}\int d^{d}x \sqrt{g}  \left(f(R) -2\Lambda\right)=\frac{A}{4} f'(R).
    \end{align}
In the last step, we use \eqref{R ds} and \eqref{Lambda}, and that the integral over the Euclidean time coordinate equals the inverse Gibbons-Hawking temperature. 

In this approach, energy is set to zero (\( E=0 \)) since in General Relativity energy follows from the boundary term and there is no boundary in de Sitter space to define it. Furthermore, using \eqref{Wq}, we can show that the energy in de Sitter space vanishes:
\begin{gather}
\label{energy de sitter}
    \beta_H E=-\beta\partial_\beta \log(Z)\bigg|_{\beta=\beta_H}= \\=\nonumber-\frac{1}{16 \pi}\int d^{d}x \sqrt{g}  \left(f \left(R^{\mathcal{M}}\right)-2 \Lambda\right) +
         \frac{1}{4}\int d^{d}x \sqrt{g}  \delta\left(\Sigma\right)  f'\left(R^{\mathcal{M}}\right) =0,
\end{gather}
where we use \eqref{Lambda} and \eqref{volum and area}.

\item \textbf{First law}\\
To write the first law of thermodynamics in the de Sitter space, we must consider how the entropy changes as a small amount of energy (e.g., a black hole) is added into this space-time. 

Let us examine how the area \( A \) of the cosmological horizon changes when an infinitesimal amount of energy \( \delta E=\delta M \) is added, where \( \delta M \) is the black hole mass. The metric of a Schwarzschild-de Sitter black hole in \( d=4 \) dimensions is:  
    \begin{align}
        ds^2=-\left(1-2 \delta M/r-r^2 H^2\right)dt^2+\frac{dr^2}{\left(1-2 \delta M/r-r^2 H^2\right)}+r^2 d\Omega^2.
    \end{align}
  
If a small energy \( \delta M \) is added to empty de Sitter space, the area of the cosmological horizon decreases as \cite{Gibbons:1976ue,Spradlin:2001pw}: 
    \begin{align}
        \delta A_H = -\frac{8 \pi }{H} \delta M.
    \end{align}
 Using the temperature defined by \eqref{T hawking} and assuming the entropy is \( 1/4 \) of the horizon area, the first law of thermodynamics holds for the cosmological horizon: 
    \begin{align} 
        \delta (-E) = T_H d S,
    \end{align}
where the black hole horizon's area is ignored since it is of the second-order in the variation of energy (\( \delta A_{BH}\sim \delta M^2 \)).

The change in energy contains the negative sign to reflec the fact that the entropy decreases as the energy inside de Sitter space increases. Indeed, as a mass leaves the cosmological horizon, it suggests that less information is accessible to the observer inside this cosmological horizon, increasing the entropy \cite{Anninos:2012qw}. 

Note, however, that there are still debates regarding the negative sign in the first law for de Sitter horizons \cite{Banihashemi:2022htw}. 

\item \textbf{Local thermodynamics}\\
If we assume that the cosmological term is generated by some matter field with an effective action: $Z=e^{-\frac{1}{16\pi}\int d^d x \sqrt{g} \ 2\Lambda}$, then the stress-energy tensor of this matter is given by: $T_{\mu\nu}=-\frac{\Lambda}{8\pi} g_{\mu\nu}$. Now, let us consider the first law of thermodynamics with non-zero pressure: 
\begin{align}
    T d S_V = d E + p d V.
\end{align}
Assuming the temperature is given by the Gibbons-Hawking relation \eqref{T hawking}, the energy density is \(\rho = \frac{\Lambda}{8\pi}\) and the pressure is \(p = -\rho\). For \( d=4 \), it follows that \cite{Padmanabhan:2002sha,Padmanabhan:2002ji}:
\begin{align}
    \frac{H}{2\pi} d S_V = d\left(\frac{4\pi}{3 H^3} \frac{3H^2}{8\pi}\right) - \frac{3 H^2}{8\pi} d\left(\frac{4\pi}{3 H^3}\right) = \frac{1}{H^2} dH.
\end{align}
Integrating the above equation gives: 
\begin{align}
\label{SVd4}
    S_V = -\frac{\pi}{H^2} = -\frac{A}{4}.
\end{align}
If we assume, as mentioned above, that the first law of thermodynamics for the cosmological horizon requires an additional negative sign, we recover the correct entropy definition \eqref{entropy f(R) general}. 

The superscript \( V \) in \eqref{SVd4} indicates that the entropy is calculated for the entire volume of the system using thermodynamic properties of the de Sitter Euclidean ground state\footnote{We refer to as the Eucledean vacuum to the gravity theory ground state under consideration, because we obtain the gravity theory in the de Sitter space via analytical continuation.}, rather than relating it solely to the horizon. This method will be explored in greater detail in the next section for the general $f(R)$ theory of gravity. 
\end{enumerate}
\section{Bulk entropy: for gravity \label{4}}
In this section we consider the cosmological constant term in general $f(R)$ theory of gravity \eqref{eq motion f(R)} as an effective action of some matter. Then using the expression of the Hubble constant in terms of the Gibbons-Hawking temperature \eqref{T hawking}, we can determine how the stress-energy tensor of such matter depends on the temperature:
\begin{align}
\label{tmunufR}
    T_{\mu \nu} = -\frac{\Lambda}{8\pi} g_{\mu\nu} = -\frac{1}{8\pi} g_{\mu\nu}\left(\frac{1}{2} f(R) - \frac{1}{d} f'(R) R \right) \Bigg|_{R=(d-1)d(2\pi T_H)^2}.
\end{align}
This approach is similar to that considered in a series of papers \cite{Volovik:2022vvi, Volovik:2023phl, Volovik:2023wty, Volovik:2024eni}, but in those works, it is assumed that the local temperature is twice that of Gibbons-Hawking. By using the expression of the stress-energy tensor in terms of hydrodynamic quantities as 
\begin{align}
    T_{\mu\nu} = (\rho + p) u_\mu u_\nu + p g_{\mu\nu},
\end{align}
where \( u_\mu = (\sqrt{|g_{00}|}, 0, 0, 0) \) is the four-velocity, we can obtain the values for the energy density \(\rho\) and pressure \(p\). Hence, we have the following equation of state:
\begin{align}
\label{eq of state}
    \rho = \frac{1}{8\pi}\Lambda(T_H) \quad \text{and} \quad p = -\frac{1}{8\pi}\Lambda(T_H),
\end{align}
with positive energy density and negative pressure, both of which are functions of the Gibbons-Hawking temperature. This equation of state describes the local thermodynamic properties of the de Sitter Euclidean vacuum. It is important to note that the geometry of the  de Sitter space  is fully determined by the Hubble constant, whereas the vacuum energy density \eqref{eq of state} is determined by the theory of gravity. For Einstein's theory of gravity, we have: 
\begin{align}
\label{lambda for de sitter}
    \rho = \frac{(d-2)(d-1)}{4} \pi T^2_H.
\end{align}
Now we are ready to compute the entropy of the de Sitter Euclidean vacuum state by using the standard laws of thermodynamics. 

The first law of thermodynamics requires that: 
\begin{align}
    \delta Q = \delta E + p \delta V,
\end{align}
where \(\delta Q\) is the infinitesimal transfer of heat to the system, \(\delta E\) is the change in the internal energy of the system, and \(p \delta V\) represents the work done by the system. 

By defining the vector current \(j^\mu = T^{\mu \nu} k_\nu\), where \(k_\nu = (g_{00}, 0, 0, ..., 0)\) is the time-like Killing vector, we can show that the vector current is conserved, \(\triangledown_\mu j^\mu = 0\). Hence, using Stokes' theorem, we can define the energy of the system: 
\begin{align}
    E = \int_{\Sigma_1} d \Sigma \sqrt{g_{d-1}} j^\mu n_\mu = \int d^{d-1} x \sqrt{g} \rho,
\end{align}
where \(n^\mu = (\sqrt{|g^{00}|}, 0, 0, ..., 0)\) is the time-like unit vector. 

The change of the internal energy is given by the change in the energy density and the volume of space: 
\begin{align}
    \delta E = \int d^{d-1} x \sqrt{g}\ \delta \rho + \delta\left(\int d^{d-1} x \sqrt{g}\right) \rho,
\end{align}
where the second term is exactly the work done by the system, with the minus sign:
\begin{align}
    \delta\left(\int d^{d-1} x \sqrt{g}\right) \rho = -\delta\left(\int d^{d-1} x \sqrt{g}\right) p = -p \delta V.
\end{align}
Thus, the infinitesimal transfer of heat is given by: 
\begin{align}
    \delta Q = \int d^{d-1} x \sqrt{g}\ \delta \rho.
\end{align}
According to the second law of thermodynamics, the change in the entropy of a system is proportional to the infinitesimal amount of heat transferred to the system divided by the temperature of the system:
\begin{align}
    \delta S = \frac{\delta Q}{T_H}.
\end{align}

Thus, the total entropy of the de Sitter Euclidean vacuum state is given by: 
\begin{align}
\label{S_V}
    S^\text{on shell}_V = \int \frac{\delta Q}{T_H} = \int \frac{d T_H}{T_H}  \int_{r_{ds} = \frac{1}{2 \pi T_H}} d^{d-1} x \sqrt{g}\ \frac{\partial \rho }{\partial T_H},
\end{align}
where the volume integral depends on the temperature, and the index \(V\) in \(S_V\) indicates that the entropy is of the bulk type. Hence, for Einstein gravity, we can obtain the following value for the entropy: 
\begin{align}
    S^\text{on shell}_V = -\frac{\pi^{\frac{d}{2}-\frac{1}{2}} H^{2-d}}{2 \Gamma\left(\frac{d-1}{2}\right)} = -\frac{A}{4},
\end{align}
up to the minus sign. Here, the key problem is that the entropy diverges in the flat space limit. This means that the entropy does not approach a constant value as the temperature approaches zero. Nevertheless, for the matter contribution, this problem is absent, as we will see further on.

Similarly, using the expression for the energy density in \(f(R)\) gravity \eqref{tmunufR}: 
\begin{align}
    \rho = \frac{1}{8\pi d}\left(   f'\left[-(d-1)d(2\pi T_H)^2\right] \left(-(d-1)d(2\pi T_H)^2\right) - \frac{d}{2} f\left[-(d-1)d(2\pi T_H)^2\right]\right),
\end{align}
we can find that the entropy of the de Sitter Euclidean vacuum state is given by: 
\begin{align}
\label{SVgrav}
    S^\text{on shell}_V = -\frac{A}{4} f'(R),
\end{align}
which exactly coincides with the Wald entropy of the Killing horizon \eqref{entropy f(R) general} up to the minus sign. Hence using the first law of thermodynamics with minus sign we obtain a coinsistent answer. As a result the definition of the bulk entropy is given by:
\begin{align}
\label{S_Vwithminussign1}
    S^\text{on shell}_V = -\int \frac{\delta Q}{T_H} = -\int \frac{d T_H}{T_H}  \int_{r_{dS} =\frac{1}{2 \pi T_H}} d^{d-1} x \sqrt{g}\ \frac{\partial \rho(T_H) }{\partial T_H}.
\end{align} 

The main point here is that \(S_V\) represents the total entropy of the volume bounded by the cosmological horizon, and that this entropy is derived from the thermodynamic properties of the Euclidean vacuum of the gravity theory, rather than being directly related to the cosmological horizon. Thus, the bulk entropy precisely follows the area law in any dimension of space and in any theory of \(f(R)\) gravity.

\section{Bulk entropy: for matter \label{5}} 
Above we have considered semiclassical contribution to the entropy, i.e. the functional integration over metrics was performed in the saddle-point approximation. (And calculation was performed in Euclidean signature). In this section we consider quantum corrections to the entropy due to matter fields. It is supposed that, in general, corrections to the entropy can be represented as: 
\begin{align}
    S_A=\frac{A}{4 G \hbar}+\alpha_0 \log\left(\frac{A}{4 G \hbar}\right)+\sum_{n}\alpha_n \left(\frac{A}{4 G \hbar}\right)^{-n},
\end{align}
where $\alpha_n$ are a theory dependent constants. We restore, for a moment, Planck's and Newton's constants. 

Now, let us consider quantum corrections to the Einstein field equations due to matter in de Sitter space: 
\begin{align}
    R_{\mu\nu}-\frac{1}{2}R g_{\mu \nu}=-\Lambda g_{\mu\nu} +8\pi \langle T_{\mu\nu} \rangle.
\end{align}
The expectation value of the stress-energy tensor of a non-minimally coupled scalar field in the Bunch-Davies state is proportional to the metric tensor \cite{Dowker,Birrell_Davies_1982,Bernard:1986vc}: 
\begin{gather}
\label{set}
    \langle T_{\mu\nu} \rangle 
    =\\= 
    \nonumber
   - g_{\mu\nu} \frac{H^4}{64\pi^2}\Bigg[\left(\frac{1}{4}+\nu ^2\right) \left(\frac{9}{4}+\nu ^2-12 \xi
   \right)\left[\psi\left(\frac{3}{2}+i\nu\right)+\psi\left(\frac{3}{2}-i\nu\right)-\log\left( \frac{m^2}{H^2}\right)\right]
    +\\+
    \nonumber
    \nu ^2 \left(\frac{4}{3}-12 \xi \right)+\xi  (72 \xi
   -19)+\frac{16}{15}   \Bigg],
\end{gather}
where: 
\begin{align}
    \nu=\sqrt{\frac{m^2}{H^2}+12 \xi -\frac{9}{4}}.
\end{align}
Hence, the self-consistent equation takes the form: 
\begin{align}
\label{selfconseqforhabble}
     -12 H^2=-4 \Lambda +8\pi \langle T^\mu_\mu \rangle.
\end{align}
As a result, for a given Hubble constant, the vacuum energy is given by: 
\begin{align}
\label{H correc}
    \rho_{\text{vac}}=\frac{\Lambda}{8\pi}= \frac{3 H^2}{8\pi}+\frac{\langle T^\mu_\mu \rangle}{4},
\end{align}
where we denote the second term as:
\begin{align}
    \rho_{\text{matter}}=-\langle T^0_0 \rangle =-\frac{\langle T^\mu_\mu \rangle }{4}.
\end{align}
Only for the Bunch-Davies state in the expanding Poincare and static patches quantum corrections lead to a redefinition of the Hubble constant without changing the geometry \cite{Akhmedov:2022uug}. In the contracting Poincare patch and in global de Sitter space-time infrared eﬀects violate the isometries and loop corrections can cause strong backreaction \cite{Akhmedov:2013vka,Akhmedov:2021rhq}.  

 As a result, the temperature is determined by the new quantum-corrected Hubble constant: \( T_H = \frac{H}{2\pi} \), where \( H \) is a solution to \eqref{selfconseqforhabble}. 

Using \eqref{S_Vwithminussign1}, we can express the total quantum-corrected entropy of the Bunch-Davies state as follows:
\begin{gather}
\label{S_V mid}
    S_V =-\int \frac{d T_H}{T_H}  \int_{r =\frac{1}{2 \pi T_H}} d^{d-1} x \sqrt{g}\ \frac{\partial \rho_{\text{vac}}(T_H) }{\partial T_H}
    =\\=\nonumber
    \frac{A}{4}+ \int_0^{T_H} \frac{d T_H}{T_H}  \int_{r=\frac{1}{2\pi T_H}} d^{3} x \sqrt{g}\ \frac{\partial \rho_{matter} }{\partial T_H},
    \end{gather}
where the first term is the standard gravity contribution, $S_V(\text{gravity})=\frac{A}{4}$. For convenience, we denote the second term as the matter contribution to the entropy, $S_V(\text{matter})$. However, matter also contributes to the first term since it is written in terms of the corrected Hubble constant, which is defined by the equation \eqref{H correc} for the given cosmological constant.

For a massive field theory one can obtain the following expression, which we will compare with the area law obtained off-shell in the next section:
\begin{gather}
S_V=\frac{A}{4}+\\ +\frac{1}{6} \zeta '(-1)-\frac {1}{72}+\frac{1}{90} \zeta '(-3)-\frac{\nu ^4}{48}-\frac{7 \nu ^2}{288}+6 \xi ^2-\frac{13 \xi }{6}-\frac{\zeta (3)}{8 \pi ^2}+\frac{2417}{11520}
   \nonumber
   +\\+
   \nonumber
  \frac{1}{720} \left(60 \nu ^2 (6 \xi -1)+90 \xi -19\right) \log \left(\frac{m^2}{4 \pi ^2 T_H^2}\right)
   +\\+
   \nonumber
   \Re e\Bigg[ 
   2 \psi ^{(-4)}\left(\frac{1}{2}+i \nu \right)
   -2 i \nu  \psi^{(-3)}\left(\frac{1}{2}+i \nu \right) -\left(\nu ^2+\frac{1}{12}\right) \psi ^{(-2)}\left(\frac{1}{2}+i \nu \right)
   +\\+
   \nonumber
   \frac{1}{12} i \left(4 \nu ^3+\nu \right) \log \Gamma\left(\frac{1}{2}+i \nu \right)
   +
   \frac{1}{192} \left(4 \nu ^2+1\right) \left(4 \nu ^2-48 \xi +9\right)\psi \left(\frac{1}{2}+i \nu \right)  \Bigg],
   \end{gather}
   where: 
   \begin{align}
       \nu=\sqrt{\frac{m^2}{4 \pi ^2 T_H^2}+12 \xi -\frac{9}{4}}.
   \end{align}
In the zero temperature limit, \(S_V(\text{matter})\) vanish, and the main contribution comes from the gravity: \(S_V(\text{gravity})= \frac{A}{4}=\pi\frac{1}{(2\pi T_H)^3}\). In the large temperature limit, the dominant contribution arises from the matter term.

For a conformal field theory (\(m=0\) and \(\xi=\frac{1}{6}\)), the trace of the stress-energy tensor is given by \cite{Bernard:1986vc}: 

\[
\langle T^\mu_\mu \rangle = -\frac{H^4}{240\pi^2}.
\]
Using this expression, we can obtain the following exact result for the entropy:
 \begin{align}
 \label{logdSentropy}
    S_V=\frac{A}{4} +\alpha_0\log \left(\frac{A}{4} M^2 \right),
\end{align}
where:
\begin{align}
    \alpha_0= -\frac{1}{180},
\end{align}
this coincides with one half of  the integrated conformal anomaly:
\begin{align}
    \alpha_0=\frac{1}{2}\int d^4 x \langle T^\mu_\mu \rangle,
\end{align}
 and \(M\) is a dimensional parameter that needs to be introduced to make the argument of the logarithm dimensionless. Such a correction term has also been discussed in the context of cosmological space in \cite{Lidsey:2008zq,MohseniSadjadi:2010nu,Maldacena:2012xp,Sharif:2014esa,Ahmed:2019lii}. The relation between $\alpha_0$ and the integrated conformal anomaly is universal for conformal field theory \cite{Solodukhin:2008dh,Aros:2010jb,Fursaev:1994te}.

Now let discuss shortly quantum corrections to the black hole entropy. Note that the following  point does not take into account interactions that lead to a significant modification of Hawking radiation due to secularly growing loop corrections \cite{Akhmedov:2015xwa} and thus should significantly change the quantum correction to the background geometry for small black holes. Nethertheless if we consider tree level theory then the logarithmic term is given by \cite{Fursaev:1994te}:
\begin{align}
    S=S_0+\alpha_0\log\left(S_0\right)+... , \quad \text{where} \quad \alpha_0= \frac{1}{90}  .
\end{align}
In this case there also should be non-logarithmic terms, since this result was obtained without considering the backreaction caused by quantum matter on the Schwarzschild metric \cite{Fursaev:1994te}. In \cite{Xiao:2021zly} the backreaction to the black hole geometry was taken into account, and the quantum correction to the black hole entropy discussed. At the same time, in de Sitter space, the result \eqref{logdSentropy} is exact for the Bunch-Davies state.  

For macroscopic black holes quantum corrections are negligibly small, whereas at the late stage of black hole evaporation quantum corrections become very important. The effective temperature of a black hole is \cite{Fursaev:1994te}: 
\begin{align}
  \label{black hole effective temperature}
    \frac{1}{T_H}=\frac{\partial S_A}{\partial M} =8 \pi  M +2\alpha_0 \frac{1}{M},
\end{align}
where we neglect subleading terms due to the smallness of \(\hbar\). Hence, for large black holes, the Hawking temperature is inversely proportional its mass \(T_H \sim \frac{1}{M}\), while for small masses, the temperature is linear to its mass \(T_H \sim M\). This means that the temperature does not blow up at the final stage, and the time of evaporation increases.

The quantum logarithmic corrections to the entropy of black holes have been derived in different contexts; \cite{Mann:1997hm,Kaul:2000kf,Das:2001ic,Modak:2008tg,Cai:2009ua,Solodukhin:2011gn,Solodukhin:2019xwx}, yielding different values for the logarithmic prefactor. For more details, see \cite{Page:2004xp,Cai:2008ys}. Moreover, the final stage of the black hole evaporation depends on the sign of \(\alpha_0\), for more details, see \cite{Fursaev:1994te, Solodukhin:2011gn}.

\section{Why does the bulk entropy of matter obeys the area law?  \label{6}}

In this section we clarify why bulk (on shell) and area (off shell) entropies should coincide, and show where there is a subtle point that requires us to check the details of calculations explicitly.

Let us expand the effective action of the matter field on a Euclidean space with a conical singularity near the Gibbons-Hawking temperature: 
\begin{align}
   \log  Z_\alpha = - W_0\left(T_H\right)-(1-\alpha)W_1 \left(T_H\right)+O\left((1-\alpha)^2\right), \quad \text{where} \quad \alpha=\frac{\beta}{\beta_H}, \quad \beta_H=\frac{1}{T_H}.
\end{align}
Then the off-shell method to calculate the entropy gives:
\begin{align}
     S^{\text{off shell}}_A=-(\alpha \partial_\alpha-1)\log Z_{\alpha} \Big|_{\alpha=1}= -W_0(T_H)-W_1(T_H).
\end{align}
The energy of the system is defined as follows:
\begin{align}
\label{derbetapath integral}
  \beta_H E =  -\beta\partial_\beta \log Z_{\alpha} \Big|_{\beta=\beta_H}=  -\partial_\alpha \log Z_{\alpha} \Big|_{\alpha=1}= -W_1(T_H).
\end{align}
The free energy obeys the standard relation:
\begin{align}
\label{statmechrel}
    F(T_H)=T_H W_0(T_H)=E(T_H)- T_H S^{\text{off shell}}_A(T_H).
\end{align}
At the same time, if we take the derivatives of the effective action with respect to the Gibbons-Hawking temperature, we obtain:
\begin{gather}
\label{free from variation}
-\beta_H\frac{d}{d \beta_H}  \log Z_{\alpha=1} = T_H\frac{d}{d T_H}  \log Z_{\alpha=1} = H \frac{d}{d H} \log Z_{\alpha=1}=\\=\nonumber 2\int d^d x  g^{\mu\nu}\frac{\delta}{\delta g_{\mu\nu}} \log Z_{\alpha=1}=-\int d^d x \sqrt{g} \langle T_{\mu}^{\mu} \rangle ^{\text{matter}}=-d \int d^d x \sqrt{g} \langle T_{0}^{0} \rangle ^{\text{matter}},
\end{gather}
where we use the fact that the change in the Gibbons-Hawking temperature corresponds to the change in the radius of the sphere (of Euclidean de Sitter space). As a result, the derivative with respect to the Gibbons-Hawking temperature is related to the variational derivative with respect to the metric. Hence, the free energy can be computed as follows:
\begin{align}
    \beta_H F(T_H)= -d \int \frac{d T_H}{T_H}\int_{r =\frac{1}{2 \pi T_H}} d^d x \sqrt{g} \rho_{\text{matter}}(T_H), \quad \text{where} \quad \rho_{\text{matter}}(T_H)=- \langle T^0_0\rangle,
\end{align}
and we take the volume integral over the sphere with the radius set by  the Gibbons-Hawking temperature or the size of the cosmological horizon. Using the expression of the energy of the system:
\begin{align}
    \beta_H E(T_H)=\int_{r =\frac{1}{2 \pi T_H}} d^d x\sqrt{g} \rho_{\text{matter}}(T_H),
\end{align}
one can obtain:
\begin{gather}
    S^\text{off shell}_A(\text{matter}) =\beta_H E(T_H)-\beta_H F(T_H)
    =\\= \nonumber
   \int d T_H \frac{\partial}{\partial T_H} \int_{r =\frac{1}{2 \pi T_H}} d^d x\sqrt{g} \rho_{\text{matter}}(T_H)+d \int \frac{d T_H}{T_H}\int_{r =\frac{1}{2 \pi T_H}} d^d x \sqrt{g} \rho_{\text{matter}}(T_H)
    =\\=
    \nonumber  \int \frac{d T_H}{T_H}  \int_{r =\frac{1}{2 \pi T_H}} d^{d-1} x \sqrt{g}\ \frac{\partial \rho_{\text{matter}}(T_H) }{\partial T_H}= S^\text{on shell}_V(\text{matter}).
\end{gather}
Hence, the equality between the two expressions for the energy  becomes evident. Nevertheless, we use at an intermediate step that derivative of the Euclidean function integral with respect to inverce temperature gives the energy of the system \eqref{derbetapath integral}. This relaion holds if there is the equality between the Euclidean functional integral and the thermal partition function:
\begin{align}
   Z_{\alpha}=Z^E =^? Z^C=\text{Tr}[e^{-\beta \hat{H}}],
\end{align}
and if the energy of the system coincides with the expectation value of the Hamiltonian: 
\begin{align}
    \langle \hat{E}\rangle=^?\langle \hat{H}\rangle=-\partial_\beta \log Z^C. 
\end{align}
Both equalities hold in the absence of Killing horizon but for space-times with Killing horizons, this relation can break down \cite{Iyer:1995kg, Frolov:1998vs,Diakonov:2023jdk} . The diffence between the energy defined by the stress-energy tensor and the canonical Hamiltonian is given by:
\begin{align}
\label{Ehxi}
    E=H -\xi Q ,
\end{align}
where ($\xi Q$) is a boundary term, which vanishes for minimally coupled field theory or in the abcense of Killing horizons, for more details see  \cite{Iyer:1995kg, Frolov:1998vs,Diakonov:2023jdk}. Moreover, as was shown in \cite{Diakonov:2023jdk}, there is a difference between the non-renormalized Euclidean functional intergal and the thermal partition function in space-times with Killing horizons for arbitryary $\beta$. Nethertheless there can be equality between them at the Gibbons-Hawking temperature, $\beta=\beta_H$.

The main point of \cite{Diakonov:2023jdk} is that for massive minimally coupled scalar field theory the Euclidean functional integral in a general space-time can be represented as a sum of three terms:
\begin{gather}
\label{logZm1}
 \log Z^E =  -  \int d^3 x \sqrt{g} g^{00} \sum_i \phi_i(x) \phi_i^*(x) \log\left(1- e^{-\beta \omega_i}\right)
 + \\+
 \nonumber
 \int d^3 x \sqrt{g} g^{00} \sum_i \int_\infty^{m^2} dm^2 \partial_{m^2} \left[\phi_i(x) \phi_i^*(x) \right] \log\left(1- e^{-\beta \omega_i}\right)
-\\-
\beta \int d^3 x \sqrt{g} \int_\infty^{m^2}dm^2 \sum_i \frac{1}{2\omega_i} \bigg[\triangle_3 \partial_{m^2}\phi_i(x) \phi_i^*(x) - \partial_{m^2}\phi_i(x) \triangle_3 
 \phi_i^*(x) \bigg]\frac{1}{e^{\beta \omega}-1}.
\nonumber 
\end{gather}
Only the first term leads to the equality between the Euclidean path integral and the thermal partition function:
\begin{align}
\label{partnoncomp}
\log \text{Tr}(e^{-\beta \hat{H}})=- \int d^3 x \sqrt{g} g^{00} \sum_i \phi_i(x) \phi_i^*(x) \log\left(1- e^{-\beta \omega_i}\right).
\end{align}
For spaces with compact spatial sections or for asymptotically flat ones, the sum of the second and third terms in \eqref{logZm1} vanishes \cite{Diakonov:2023jdk}, and the Euclidean functional integral $Z^E$ becomes equal to the thermal partition function of the canonical ensemble $Z^C$. However, if spatial sections are bounded by Killing horizons, the situation is quite different. In such a scenario, the single-particle energy spectrum of the theory does not depend on the mass, i.e., $\omega \neq \omega(m)$. This leads to that the first and second terms in \eqref{logZm1} canceling out each other, leaving the Euclidean functional integral defined solely by the third term:
\begin{gather}
\label{logZm21}
 \log Z^E = \\= \nonumber -
\beta \int d^3 x \sqrt{g} \int_\infty^{m^2} d m^2  \sum_i \frac{1}{2\omega_i}   \bigg[\triangle_3 \partial_{m^2}\phi_i(x)  \phi_i ^*(x) -  \partial_{m^2}\phi_i(x)  \triangle_3 
 \phi_i ^*(x) \bigg] \frac{1}{e^{\beta \omega_i}-1}.
\nonumber 
\end{gather}
In this case, the Euclidean functional integral reduces to a volume integral of a density that is a total derivative. As a result, the Euclidean functional integral on space-times with Killing horizons can be defined as a volume integral of some density or as an integral over the area of the Killing horizon:
\begin{align}
\label{A=V}
    \log Z^E = \int d^3 x \sqrt{g} ... = \int d A ...,
\end{align}
which implies that there should be a connection between the bulk and the boundary (horizon) contributions to the entropy.

If we change the order of integration over the volume of space and over the momentum in \eqref{logZm21} (we do that formally since this procedure is not well defined due to the divergence of the volume integral \cite{Diakonov:2023jdk}), we obtain that the Euclidean functional integral can be represented in terms of the difference between two single-particle densities of states, $\rho$ and $\rho_{\text{ref}}$, where the reference density of states is chosen on a Rindler-like region. For more details, see \cite{Law:2022zdq,Grewal:2022hlo,Law:2023ohq,Law:2025ktz}:
\begin{gather}
 \log Z^E = -  \sum_i     \left(\rho - \rho_{\text{ref}}\right) \log\left(1 - e^{-\beta \omega_i}\right).
\end{gather}
Thus, following the works \cite{Law:2022zdq,Grewal:2022hlo,Law:2023ohq,Law:2025ktz} the difference between the Euclidean functional integral and the thermal partition function can be represented as:
\begin{align}
    Z^E = \frac{\text{Tr} \left[ e^{-\beta \hat{H}}\right]}{\text{Tr}_{\text{Rindler}} \left[ e^{-\beta \hat{H}}\right]},
\end{align}
where $\text{Tr}_{\text{Rindler}}$ is the reference canonical partition function computed in the Rindler-like region \cite{Law:2022zdq}. 

Hence  the point of the precence of some discrepacy between the Euclidean functional integral and the thermal partition function should be verified explicitly for properly renormalized effective action. 

Below, we will explicitly compute the Euclidean functional integral in de Sitter space. And show that, after proper renormalization, there is indeed equality between the Euclidean functional integral and the thermal partition function for minimally coupled scalar field theory at Gibbons-Hawking temperatute. For the case of a non-minimal coupling the difference is still present. However, it seems that for an arbitrary temperature, the Euclidean functional integral is not equal to the canonical partition function:
\begin{align}
    \log Z^E \ne \log Z^C , \quad \text{for} \quad \beta \ne \beta_H.
\end{align}

At the end of this section, let us remark that it is not clear why both the on-shell and off-shell methods yield the same result for the entropy of the pure gravitational action, as the relation \eqref{free from variation} does not hold:
\begin{align}
  \frac{\delta}{\delta g_{\mu\nu}} \log Z(\text{gravity}) = 0,
\end{align}
and there is also a problem with of the definition of the energy since de Sitter space is not asymptotically flat: as discussed above, the total energy $E$ should be equal to zero in de Sitter space. Nevertheless, we define the energy as a volume integral over the energy density, arising from the contribution of the cosmological term, and obtain the precise equality \eqref{SVgrav}. The only possibility for the explanation that we see at this point is that if we understand the presence of the cosmological term as being generated by some matter field with an effective action: $Z=e^{-\frac{1}{16\pi}\int d^d x \sqrt{g} \ 2\Lambda}$, then the energy of this matter is given by: 
\begin{align}
    -\partial_\beta \log Z\Bigg|_{\beta=\beta_H}= \frac{\Lambda}{8\pi } V_{d-1}.
\end{align}

\section{The area law: matter \label{7}}
In this section we compute quantum corrections to the entropy using the off-shell method. To do that, we consider the Euclidean functional integral on the Euclidean static de Sitter space with generic period $\beta$ of Euclidean time, i.e., the functional integral on the sphere with conical singularities. 

In the first step we discuss the general renormalization procedure of the effective action on a space with conical singularities. Then, we show that after the proper renormalization the entropy computed off-shell indeed coincides with the entropy computed on-shell.

\subsection{Renormalization procedure}
It is known that entanglement entropy diverges and is proportional to the area of the horizon surface. If we consider the Euclidean functional integral on a conical manifold with a generic period in imaginary time ($\beta$ -- the inverse temperature), then the leading divergent contribution to the free energy is similar to that in Rindler coordinates of the flat space-time \cite{Kabat:1995eq}: $
\beta F \sim - A \frac{1}{\beta}\left(1-\frac{\beta^2}{\beta_H^2}\right) \frac{1}{\epsilon^2}$. In this case, the entropy diverges, and its contribution depends on the temperature: $S \sim A\frac{1}{\beta} \frac{1}{\epsilon^2}$. Hence, for arbitrary $\beta$, it is not clear how to renormalize the divergent contribution, since it depends on the state of the system. Nevertheless, if the temperature is equal to the Gibbons-Hawking one, which can be considered as a geometrical characteristic of space, then one can subtract this divergence by renormalizing Newton’s constant $G$ \cite{Susskind:1994sm}. Subleading divergences can also be renormalized by redefining the higher curvature couplings in the gravitational action \cite{Solodukhin:1994yz, Fursaev:1994ea, Solodukhin:1995ak}. In this section we demonstrate this explicitly, following the work \cite{Solodukhin:1995ak}.

The total effective action contains two terms: the bare gravitational action and the divergent effective action of the matter field:
\begin{align}
    W^{ren} = W^{bare}_{gr} + W^{div}_{matter}.
\end{align}
The bare gravitational action in four dimensions is expressed as:
\begin{gather}
    W^{bare}_{gr} = \\= \nonumber\int_{\mathcal{M}_\alpha} d^4 x \sqrt{g} \Bigg(-\frac{1}{16 \pi G_B}\big(R^{\mathcal{M}_\alpha} - 2 \Lambda_B\big)
    + c_{1}^B R^{\mathcal{M}_\alpha} R^{\mathcal{M}_\alpha} + c_{2}^B R^{\mathcal{M}_\alpha}_{\mu\nu} R^{\mathcal{M}_\alpha,\mu\nu} + c_{3}^B R^{\mathcal{M}_\alpha}_{\mu\nu\rho\sigma} R^{\mathcal{M}_\alpha,\mu\nu\rho\sigma} \Bigg),
\end{gather}
where $G_B$, $\Lambda_B$, $c_1^B$, $c_2^B$, and $c_3^B$ are the bare coupling constants.

A Euclidean space with the conical singularity at the horizon ($\Sigma$) can be represented as  \( \mathcal{M}_\alpha \approx \mathcal{C}_\alpha \times \Sigma \), where $\mathcal{C}_\alpha$ is the two-dimensional conical manifold. Then in the limit \( \alpha \to 1 \), the Riemannian tensor contains delta-like surface contributions near the horizon \cite{Fursaev:1995ef, Solodukhin:2011gn}:
\begin{align}
    \lim_{\alpha \to 1} R^{\mathcal{M}_\alpha}_{\mu\nu\rho\sigma} = R^{\mathcal{M}}_{\mu\nu\rho\sigma} + 2\pi(1-\alpha) \big(n_\mu^1 n_\rho^2 - n_\mu^2 n_\rho^1\big)\big(n_\nu^1 n_\sigma^2 - n_\nu^2 n_\sigma^1\big) \delta(\Sigma) + O\big((1-\alpha)^2\big),
\end{align}
where \( n^{1,2} \) are two orthonormal vectors, which are orthogonal to the horizon surface \( \Sigma \). Then, the bare gravitational action up to the first order in \( \alpha \to 1 \) is given by:
\begin{gather}
    W^{bare}_{gr} = \alpha \int_\mathcal{M} d^4 x \sqrt{g} \Bigg(-\frac{1}{16 \pi G_B}\big(R^{\mathcal{M}} - 2 \Lambda_B\big) + c_{1}^B R^{\mathcal{M}} R^{\mathcal{M}} + c_{2}^B R^{\mathcal{M}}_{\mu\nu} R^{\mathcal{M},\mu\nu} + c_{3}^B R^{\mathcal{M}}_{\mu\nu\rho\sigma} R^{\mathcal{M},\mu\nu\rho\sigma} \Bigg) + \\ \nonumber 
    + 4\pi (1-\alpha) \int d\Sigma \Bigg(-\frac{1}{16 \pi G_B} + 2 c_{1}^B R^{\mathcal{M}} + c_{2}^B R^{\mathcal{M}}_{\mu\nu} n^\mu_i n^\nu_i + 2 c_{3}^B R^{\mathcal{M}}_{\mu\nu\rho\sigma} n^\mu_i n^\rho_i n^\nu_j n^\sigma_j \Bigg) + \\ \nonumber
    + O\big((1-\alpha)^2\big).
\end{gather}

Furthermore, the Euclidean path integral for the non-minimally coupled scalar field is:
\begin{align}
    Z^{div}_{matter} = \int D\phi \ e^{-\frac{1}{2} \int_{\mathcal{M}_\alpha} d^4 x \sqrt{g} \ \phi \big(-\Box + m^2 + \xi R^{\mathcal{M}_\alpha}\big)\phi}.
\end{align}
Here, the key point is that we must take into account that the Ricci scalar contains delta-like singularity on the horizon surface: 

\begin{align}
\label{Rdelta}
R^{\mathcal{M}_\alpha} = R^{\mathcal{M}} + 4\pi(1-\alpha)\delta(\Sigma) + O\big((1-\alpha)^2\big).    
\end{align}
We can expand the terms containing the delta function up to the first order in $(1-\alpha)$:
\begin{align}
\label{logZdivmatter}
    \log Z^{div}_{matter} = \log \bar{Z}^{div}_{matter} - 2\pi\xi (1-\alpha) \int d\Sigma \langle \phi^2 \rangle^{div}_{\bar{Z}},
\end{align}
where $\bar{Z}_{matter}$ is the Euclidean functional integral on the conical manifold but without delta-like singularity contribution in the action:
\begin{align}
\label{logZbar}
    \bar{Z}^{div}_{matter} = \int D\phi \ e^{-\frac{1}{2} \int_{\mathcal{M}_\alpha} d^4 x \sqrt{g} \ \phi \big(-\Box + m^2 + \xi R^{\mathcal{M}}\big)\phi}.
\end{align}
The expectation value of $\phi^2$ is given by:
\begin{align}
    \langle \phi^2 \rangle^{div}_{\bar{Z}} = \frac{1}{\bar{Z}_{matter}} \int D\phi \ e^{-\frac{1}{2} \int_{\mathcal{M}_\alpha} d^4 x \sqrt{g} \ \phi \big(-\Box + m^2 + \xi R^{\mathcal{M}}\big)\phi} \phi^2.
\end{align}
To calculate $\bar{Z}_{matter}^{div}$ and $\langle \phi^2\rangle_{\bar{Z}}^{div}$ we can use the heat kernel expansions \cite{Fursaev:1995ef}: 
\begin{align}
    \log\bar{Z}_{matter}^{div}= \frac{1}{2} \int_{\epsilon^2}^\infty \frac{ds }{s}  e^{-s m^2}\int_{\mathcal{M}_\alpha} d^4 x \sqrt{g} \   \bar{K}_{\mathcal{M}_\alpha}(s, x, x),
\end{align}
and 
\begin{align}
    \langle \phi^2\rangle_{\bar{Z}}^{div} = \int_{\epsilon^2}^\infty ds  e^{-s m^2 } \bar{K}_{\mathcal{M}_\alpha}(s, x, x).
\end{align}

The heat kernel on a manifold with a conical singularity is given by: 
\begin{align}
    \bar{K}_{\mathcal{M}_\alpha}(s, x, x) =\frac{1}{(4\pi s)^{\frac{d}{2}}}\sum_n \bar{a}_n(x)s^n,
\end{align}
with the coefficients expressed as:
\begin{align}
    \bar{a}_n(x) = a_n^{st}(x) + \left(1 - \alpha\right)a_n^\alpha(x)\delta(\Sigma) + O\left((1-\alpha)^2\right),
\end{align}
where the coefficients $a_n^{st}(x)$ are the standard heat kernel coefficients expressed as local functions of the curvature tensor \cite{Birrell_Davies_1982}:
\begin{gather}
    a_0^{st}(x)=1, \quad a_1^{st}(x)=\left(\frac{1}{6}-\xi\right)  R^\mathcal{M}, \\ \nonumber 
    a_2^{st}(x)= \frac{1}{180} R^{\mathcal{M}}_{\mu\nu\rho\sigma}R^{\mathcal{M},\mu\nu\rho\sigma} - \frac{1}{180} R^{\mathcal{M}}_{\mu\nu}R^{\mathcal{M},\mu\nu} - \frac{1}{6}\left(\frac{1}{5}-\xi\right)\Box R^{\mathcal{M}} + \frac{1}{2} \left(\frac{1}{6}-\xi\right)^2  R^\mathcal{M} R^\mathcal{M}.
\end{gather}
Additional terms $a_n^\alpha(x)$ arise due to the presence of the conical singularity \cite{Fursaev:1995ef}:
\begin{gather}
    a_0^\alpha(x)=0, \quad a_1^\alpha(x)=4\pi, \\ \nonumber 
    a_2^\alpha(x)=\frac{2}{3}\pi \left(\frac{1}{6}-\xi\right) R^\mathcal{M} - \frac{1}{45}\pi \left(R^{\mathcal{M}}_{\mu\nu}n^\mu_i n^\nu_i - 2 R^{\mathcal{M}}_{\mu\nu\rho\sigma} n^\mu_i n^\rho_i n^\nu_j n^\sigma_j\right).
\end{gather}
Hence, the Euclidean functional integral $Z^{div}_{matter}$ has the following expansion:
\begin{align}
\label{logZdiv}
    \log Z^{div}_{matter}= \frac{1}{2} \int_{\epsilon^2}^\infty \frac{ds}{s}  e^{-s m^2} \int_{\mathcal{M}_\alpha} d^4 x \sqrt{g} \   K_{\mathcal{M}_\alpha}(s, x, x),
\end{align}
where
\begin{align}
\label{heat kernel exp}
  \int_{\mathcal{M}_\alpha} d^4 x \sqrt{g} \   K_{\mathcal{M}_\alpha}(s, x, x) =\frac{1}{(4\pi s)^2}\sum_n a_n s^n,
\end{align}
and
\begin{gather}
    a_n = \int_{\mathcal{M}_\alpha} d^4 x \sqrt{g} \ \bar{a}_n(x) - 4\pi \xi (1-\alpha)\int d \Sigma \ \bar{a}_{n-1}(x)
    =\\=
    \nonumber
    \alpha \int_{\mathcal{M}} d^4 x \sqrt{g} \ a^{st}_n(x)
    +
    (1-\alpha)\int d \Sigma \ \left(a_n^\alpha(x)
    -
    4\pi \xi  \ a^{st}_{n-1}(x) \right).
\end{gather}
Thus, it can be seen that up to the zeroth and first orders in $(1-\alpha)$ all divergences are completely removed by the standard renormalization of the gravitational coupling constants:
\begin{gather}
\label{Wrrenselfconsi}
   W_{ren} = W_{gr} + W_{matter}
   = \\ =\nonumber 
   \alpha\int_\mathcal{M} d^4 x \sqrt{g} \left(-\frac{1}{16 \pi G_{ren}}\left(R^{\mathcal{M}} - 2 \Lambda_{ren}\right) + c_{1}^{ren} R^{\mathcal{M}} R^{\mathcal{M}} + c_{2}^{ren} R^{\mathcal{M}}_{\mu\nu}R^{\mathcal{M},\mu\nu} + c_{3}^{ren} R^{\mathcal{M}}_{\mu\nu\rho\sigma}R^{\mathcal{M},\mu\nu\rho\sigma}\right) +
    \\+ \nonumber 
    4\pi (1-\alpha)\int d \Sigma \left(-\frac{1}{16 \pi G_{ren}} + 2 c_{1}^{ren} R^{\mathcal{M}} + c_{2}^{ren} R^{\mathcal{M}}_{\mu\nu}n^\mu_i n^\nu_i + 2 c_{3}^{ren} R^{\mathcal{M}}_{\mu\nu\rho\sigma} n^\mu_i n^\rho_i n^\nu_j n^\sigma_j\right)+\\+\nonumber
    O\left((1-\alpha)^2\right)+\text{finite terms}.
\end{gather}
The renormalized constants are given by:
\begin{gather}
\label{renormal}
     \frac{1}{G_{ren}} = \frac{1}{G_B} + \frac{(6\xi - 1)}{12 \pi \epsilon^2} + \frac{m^2 (6\xi - 1)}{6 \pi } \log (\epsilon m),
     \\
     \nonumber
     c_1^{ren} = c_1^B - \frac{(1-6 \xi )^2 \log (e m)}{1152 \pi ^2},
     \\
     \nonumber
     c_2^{ren} = c_2^B + \frac{1}{2880 \pi ^2}\log (\epsilon m), 
     \\
     \nonumber
     c_3^{ren} = c_3^B - \frac{1}{2880 \pi ^2}\log (\epsilon m),
     \\ 
     \nonumber
     \frac{\Lambda_{ren}}{G_{ren}} =  \frac{\Lambda_{B}}{G_{B}} - \frac{1}{8 e^4 \pi} + \frac{m^2}{4 e^2 \pi} + \frac{m^4 \log (e m)}{4 \pi}.
\end{gather}
Thus, it is sufficient to utilize the standard renormalization procedure up to the second order in \( (1 - \alpha)^2 \). Note that if we do not take into account the delta-like contribution to the Ricci scalar on the conical manifold \eqref{Rdelta}, which provides the second term in \eqref{logZdivmatter}, then there will not be a self-consistent renormalization procedure at the zeroth and first orders in \( (1 - \alpha) \) in \eqref{Wrrenselfconsi}. At the second order the heat kernel coefficients are not well-defined \cite{Fursaev:1995ef}, as their contributions behave like \( \delta^2(\Sigma) \). Hence, it is not clear whether the standard renormalization procedure is well-defined at higher orders. Nevertheless, this contribution is not necessary to compute the entropy or the energy at $\beta=\beta_H$, $\alpha=1$.

Note also that if consider a thermal gas with the thermal density matrix with an arbitrary temperature, different from the canonical one $\beta\ne \beta_H$, then correlation functions do not posses Hadamard properties on Killing horizons and the back-reaction on the background geometry is strong (see, for example: \cite{Candelas:1980zt, Fulling:1977zs,Akhmedov:2019esv, Akhmedov:2020ryq,Akhmedov:2020qxd, Anempodistov:2020oki, Akhmedov:2021cwh, Bazarov:2021rrb,Akhmedov:2022qpu}). Hence, the method of computation of the partition function for generic $\beta$ does not take into account the fact that if we consider $\beta \ne \beta_H$, then the geometry of space-time may change significantly. Moreover, if we consider interacting field theory and if the temperature is less than the canonical one, then the theory becomes unstable due to the presence of a tachyon excitation \cite{Diakonov:2023hzg}. Hence, the derivatives of the Euclidean functional integral with respect to $\beta$ are not well-defined if $\beta < \beta_H$ in interacting field theories. We do not address all these points in this paper. Nevertheless, as we will see below, the entropy computed off-shell indeed coincides with the entropy computed on-shell.

\subsection{Explicit calculation}
The Euclidean functional integral \eqref{logZbar} in de Sitter space for the scalar field theory with an arbitrary temperature has been computed in several papers: using the zeta function \cite{Fursaev:1993hm}, using Harish-Chandra characters \cite{Anninos:2020hfj}, and using the Green function at coinciding points \cite{Akhmedov:2021cwh}. In these works, the authors compute $\log \bar{Z}_{matter}^{div}$, which differs from $\log Z_{matter}^{div}$ due to the boundary term $\sim \xi \int d\Sigma \langle \phi^2 \rangle^{div}_{\bar{Z}}$. This boundary term is necessary to define the proper renormalization procedure \eqref{renormal} for the non-minimally coupled theory, $\xi \ne 0$.

The divergence in \eqref{logZdiv} arises from the few first terms of the heat kernel expansion \eqref{heat kernel exp}, namely $a_0, a_1$, and $a_2$. Therefore, we can use the minimal DeWitt-Schwinger subtraction scheme to renormalize the effective action:
\begin{gather}
\label{logZrenMatterminusdiv}
    \log Z^{ren}_{matter} = \\= \nonumber
    \log \bar{Z}^{div}_{matter} - 2\pi\xi (1-\alpha) \int d\Sigma \langle \phi^2 \rangle_{\bar{Z}} - \frac{1}{2} \int_{\epsilon^2}^\infty \frac{ds}{s}   \frac{e^{-s m^2}}{(4\pi s)^2}\left(a_0 + a_1 s + a_2 s^2\right).
\end{gather}
Furthermore we use the zeta function regularization instead of the heat kernel one, since the expansion of $\log \bar{Z}^{div}_{matter}$ in terms of $(1-\alpha)$ has been computed using the zeta function techniques in \cite{Fursaev:1993hm}:
\begin{align}
    \log \bar{Z}^{div}_{matter} = \frac{1}{2}\left(\zeta'(0,\beta) + \log\left(\frac{\mu^2}{H^2}\right) \zeta(0,\beta)\right),
\end{align}
where:
\begin{gather}
    \zeta(0,\beta) 
    \approx \\ 
    \approx  \nonumber
    \left[ \frac{\nu^4}{12} + \frac{\nu^2}{24} - \frac{17}{2880} \right] 
    + 
    \left(\frac{\beta}{\beta_H} - 1\right)\left[\nu^4 + \frac{4}{24} \nu^2 + \frac{3}{64}\right]
\end{gather}
and
\begin{gather}
    \zeta'(0,\beta) 
    \approx \\ 
    \approx \nonumber
    \left[-\frac{1}{3}\left(\int_{\frac{1}{2}}^{\frac{1}{2}+i\nu} + \int_{\frac{1}{2}}^{\frac{1}{2}-i\nu}\right) u\left(u - \frac{1}{2}\right)\left(u - 1\right) \psi(u) du + \frac{\nu^4}{12} - \frac{\nu^2}{72} + \frac{2}{3}\left(\zeta'\left(-3,\frac{3}{2}\right) - \frac{1}{4}\zeta'\left(-1,\frac{3}{2}\right)\right)\right] 
    - \\ - \nonumber
    \left(\frac{\beta}{\beta_H} - 1\right)\left[-\frac{41}{144}\nu^2 - \frac{1}{8}\nu^4 - \frac{973}{5760} + \frac{1}{192}\left(16\nu^4 + 40 \nu^2 + 9\right)\left(\psi\left(\frac{3}{2}+i\nu\right) + \psi\left(\frac{3}{2}-i\nu\right)\right)\right].
\end{gather}
To evaluate the third term in \eqref{logZrenMatterminusdiv}, we need to make the following replacement:
\begin{align}
    \int_{\epsilon^2}^\infty \frac{ds }{s}  \frac{e^{-s m^2}}{(4\pi s)^2}\left(a_0 + a_1 s + a_2 s^2\right) \to \frac{d}{dz}\bigg[\frac{\mu^{2z}}{\Gamma(z)}  \int_{0}^\infty \frac{ds }{s} s^z  \frac{e^{-s m^2}}{(4\pi s)^2}\left(a_0 + a_1 s + a_2 s^2\right) \bigg] \Bigg|_{z = 0}.
\end{align}
The second term in \eqref{logZrenMatterminusdiv} can be evaluated as follows:
\begin{gather}
    \langle \phi^2 \rangle_{\bar{Z}} = - \frac{1}{V_4 \bar{R}} \partial_\xi \log Z^{ren}_{matter}\Bigg|_{\beta = \beta_H} 
    = \nonumber
    - \frac{1}{V_4 \bar{R}} \partial_\xi \Bigg[\frac{1}{2}\left(\zeta'(0,\beta_H) + \log\left(\frac{\mu^2}{H^2}\right) \zeta(0,\beta_H)\right) - \\ 
    - \nonumber \frac{1}{2} \frac{d}{dz}\bigg[\frac{\mu^{2z}}{\Gamma(z)}  \int_{0}^\infty \frac{ds }{s} s^z  \frac{e^{-s m^2}}{(4\pi s)^2}\left(a_0^{st} + a_1^{st} s + a_2^{st} s^2\right) \bigg] \Bigg|_{z=0}\Bigg].
\end{gather}
As a result, one can obtain the renormalized effective action that does not depend on the scale parameter $\mu$:
\begin{gather}
\label{renlogZ}
    \log Z^{ren}_{matter} = -W_0\left(T_H\right) - (1-\alpha)W_1\left(T_H\right),
\end{gather}
where:
\begin{gather}
W_0\left(T_H\right) = \frac{\left(17 - 120(2 \nu^4 + \nu^2)\right)}{5760} \log\left(\frac{m^2}{H^2}\right) - \\  \nonumber
-\frac{1}{6} \zeta '(-1)+\frac {1}{72} - \frac{1}{3} \zeta'(-3) + \frac{1}{288} \nu^2\left(6 \nu^2 + 35\right) - 3 \xi^2 - \frac{1}{8}\left(4 \nu^2 - 7\right) \xi + \frac{\zeta(3)}{8\pi^2} - \frac{55}{768} 
- \\ - \nonumber
\Re\Bigg[
\frac{1}{12} i\left(4 \nu^3 + \nu\right) \log\Gamma\left(i \nu + \frac{1}{2}\right) + \\ +
\nonumber
\frac{1}{24}\left(12 \nu^2 + 1\right) \psi^{(-2)}\left(i \nu + \frac{1}{2}\right) - \psi^{(-4)}\left(i \nu + \frac{1}{2}\right) + i \nu \psi^{(-3)}\left(i \nu + \frac{1}{2}\right)\Bigg]
\end{gather}
and 
\begin{gather}
W_1(T_H) 
= \\ =
\nonumber \frac{1}{384}\left(4 \nu^2 + 1\right)\left(4 \nu^2 - 48 \xi + 9\right) \log\left(\frac{m^2}{H^2}\right) + \frac{20 \nu^2 (36 \xi - 7) + 60(31 - 72 \xi) \xi - 199}{1440} 
- \\ - \nonumber
\frac{1}{384}\left(4 \nu^2 + 1\right)\left(\psi\left(\frac{1}{2} - i \nu\right) + \psi\left(i \nu + \frac{1}{2}\right)\right)\left(4 \nu^2 - 48 \xi + 9\right).
\end{gather}
The energy of the system is defined as the variation of the effective action \eqref{free from variation}:
\begin{align}
    \langle \hat{E}\rangle = \frac{H^2}{8\pi} \frac{d}{d H}\Big(\log Z^{ren}_{matter}\Big|_{\alpha=1}\Big).
\end{align}
This expression coincides with:
\begin{align}
    \langle \hat{E}\rangle = -\int d^3 x \sqrt{g} \langle \hat{T}^0_0\rangle,
\end{align}
where the stress-energy tensor is given by \eqref{set}, which has been obtained, for example, in \cite{Dowker, Birrell_Davies_1982, Bernard:1986vc}.

One can then show that taking the derivatives with respect to the inverse temperature of the renormalized effective action \eqref{renlogZ} yields exactly the energy of the system:
\begin{align}
\label{statmech rel do not sat}
\boxed{-\Big(\partial_\beta \log Z^{ren}_{matter}\Big) \Bigg|_{\alpha=1} = \langle \hat{E}\rangle}.
\end{align}
This is the main relation of this section. From the discussion in the Section \ref{6} it is clear that if the derivatives with respect to the inverse temperature of the renormalized effective action give exactly the energy of the system, as in \eqref{statmech rel do not sat}, then there should be the equality between the off-shell and on-shell methods. Indeed, the explicit calculation shows that \( S^{\text{off shell}}_A = S^{\text{on shell}}_V \). Thus, we demonstrate that the bulk entropy computed on-shell is equal to the properly renormalized entanglement entropy computed off-shell.

At first glance, the relation \eqref{statmech rel do not sat} seems obvious; nevertheless, due to the difference between the energy defined by the stress-energy tensor and the canonical Hamiltonian \cite{Iyer:1995kg, Frolov:1998vs}:
\begin{align}
\label{Ehxi}
    \langle \hat{E} \rangle = \langle \hat{H} \rangle - \xi \langle \hat{Q}\rangle,
\end{align}
it is evident that the statistical mechanics relation is not fulfilled:
\begin{align}
\label{Z=E+Q}
    -\Big(\partial_\beta \log Z^{ren}_{matter}\Big) \Bigg|_{\alpha=1} = \langle \hat{H} \rangle - \xi \langle \hat{Q} \rangle \ne \langle \hat{H}\rangle.
\end{align}
In our case, the expectation value of the boundary term \( Q \) is given by:
\begin{align}
    \langle \hat{Q}\rangle = H \int_{r=\frac{1}{H}} d \Sigma \langle \phi^2\rangle_{\bar{Z}}.
\end{align}
If we consider the renormalized effective action \eqref{logZrenMatterminusdiv}:
\begin{align}
\label{re re}
     \log Z^{ren}_{matter} = 
    \log \bar{Z}^{ren}_{matter} - 2\pi\xi (1-\alpha) \int d\Sigma \langle \phi^2 \rangle_{\bar{Z}},
\end{align}
and take the derivatives with respect to the inverse temperature, we obtain the following relation:
\begin{align}
\label{Z=E+Q 2}
    -\Big(\partial_\beta \log Z^{ren}_{matter}\Big) \Bigg|_{\alpha=1} = -\Big(\partial_\beta \log \bar{Z}^{ren}_{matter}\Big) \Bigg|_{\alpha=1} - \xi \langle \hat{Q}\rangle.
\end{align}
Thus from \eqref{Z=E+Q} and \eqref{Z=E+Q 2} , \( \bar{Z} \) is equal to the thermal partition function since it satisfies the statistical mechanics relation:
\begin{align}
    -\Big(\partial_\beta \log \bar{Z}^{ren}_{matter}\Big) \Bigg|_{\alpha=1} = \langle \hat{H}\rangle = \frac{\text{Tr}\left[e^{-\beta_H \hat{H}} \hat{H}\right]}{\text{Tr}\left[e^{-\beta_H \hat{H}} \right]}.
\end{align}
At the same time, the free energies that arise from \( Z_{\alpha=1}\) and \( \bar{Z}_{\alpha=1} \) are equal to each other when $\beta=\beta_H$ or $\alpha=1$ \eqref{re re}. Then the entropy takes the form:
\begin{align}
    S^{\text{off shell}}_A = -\left(\beta \partial_\beta - 1\right) \log Z^{ren}_{matter} \Bigg|_{\beta = \beta_H} = \bar{S} - \xi \beta_H \langle \hat{Q}\rangle.
\end{align}
Here, \( \bar{S} \) is the standard definition of the entropy: \( \bar{S} = \beta_H \langle \hat{H}\rangle - \beta_H F \), which arises from the fact that the Euclidean functional integral \( \bar{Z} \) is equal to the partition function \( \bar{Z} = \text{Tr}\left[e^{-\beta \hat{H}}\right] \). For more details about the boundary term \( Q \) and the problem of its statistical mechanical interpretation, see \cite{Fursaev:1998hr, Solodukhin:2011gn, Fursaev:2022goi}. Here we merely point out that in the off-shell method, the origin of \( Q \) is traced to the delta-function-like term in the scalar curvature, and this term is needed to properly subtract UV divergences \eqref{renormal}. From the on-shell perspective, this term arises simply due to the difference between the energy defined by the stress-energy tensor and the canonical Hamiltonian.

   \section{Conclusions}
In this note, we demonstrate that the generalized de Sitter entropy cannot be solely viewed as just a boundary term; rather, it can be understood as a bulk contribution. It is important to note that, in the on-shell method, the source of the energy density is attributed to a cosmological constant term. Consequently, it remains unclear whether we can generalize the statement of equality between the bulk and the boundary enropies to other space-times, both with or without cosmological terms. From the eq.\eqref{A=V} it follows that the non-renormalized relation for the entropy in general space-times with Killing horizons can be interpreted as arising from either bulk or boundary contributions, expressed as \( S = \int dV \ldots = \int dA \ldots \). However, it is not clear to us how to demonstrate this explicitly, similarly to the on-shell and off-shell approaches in de Sitter space.

\section*{Acknowledgments}
I would like to thank E.T.Akhmedov for valuable discussions, sharing his ideas and correcting the text. This work was supported by the grant from the Foundation for the Advancement of Theoretical Physics and Mathematics ``BASIS'', and the state assignment of the Institute for Information Transmission Problems of RAS.

\newpage 
\bibliographystyle{unsrturl}
\bibliography{bibliography.bib}
\end{document}